\documentclass[letterpaper, preprint, paper,11pt]{AAS}	

\usepackage{bm}
\usepackage{amsmath}
\usepackage[colorlinks=true, pdfstartview=FitV, linkcolor=black, citecolor= black, urlcolor= black]{hyperref}
\usepackage{overcite}
\usepackage{footnpag}			      	

\usepackage{xcolor}

\usepackage{caption}
\usepackage{subcaption}

\usepackage{float}

\PaperNumber{23-296}

\begin{document}

\title{A CONTROL FRAMEWORK FOR CUBESAT RENDEZVOUS AND PROXIMITY OPERATIONS USING ELECTRIC PROPULSION}

\author{Bo-Chuan Lin\thanks{Master of Science, Aeronautics and Astronautics, National Cheng Kung University, p46084485@gs.ncku.edu.tw.},  
Chun-Wei Kong\thanks{PhD Student, Aerospace Engineering Sciences, University of Colorado Boulder, Chun-Wei.Kong@colorado.edu.},
Simone Semeraro\thanks{PhD Student, Aerospace Engineering, Purdue University, simonse@umich.edu.},
\ and Jay W. McMahon\thanks{Associate Professor, Ann and H.J. Smead Department of Aerospace Engineering Sciences, jay.mcmahon@colorado.edu.}
}

\maketitle{}

\begin{abstract}
A control framework is presented to solve the rendezvous and proximity operations (RPO) problem of the EP-Gemini mission. In this mission, a CubeSat chaser is controlled to approach and circumnavigate the other uncooperative CubeSat target. Such a problem is challenging because the chaser operates on a single electric propulsion thruster, for which coupling between attitude control and thrust vector, and charging of the electric propulsion system must be taken into consideration. In addition, the access to relative states in real time is not achievable due to the onboard hardware constraints of the two CubeSats. The developed control framework addresses these limitations by applying four modularized maneuver blocks to correct the chaser's mean orbit elements in sequence. The control framework is based on a relative motion called safety ellipse to ensure a low collision risk.
The complete EP-Gemini mission is demonstrated by the implementation of the proposed control framework in a numerical simulation that includes high order perturbations for low Earth orbit. The simulation result shows that a safety ellipse is established after a 41-day RPO maneuver, which consumes 44$\%$ of the total fuel in terms of $\Delta V$. The resulting 3-dimensional safety ellipse circumnavigates the target with an approximate dimension of 14 km $\times$ 27 km $\times$ 8 km. 
\end{abstract}

\section{Introduction}
The demand for rendezvous and proximity operations (RPO) for CubeSats has grown recently. In 2020, AeroCube-10 1.5U CubeSats performed an RPO from the initial in-track separation of 1600 km to a passively safe formation of 50 meters radius \cite{gangestad2021sat}. In May 2022, NASA's CubeSat Proximity Operations Demonstration (CPOD) launched two 3U CubeSats to low Earth orbit (LEO), which will validate low-power RPO and docking for nanosatellites \cite{roscoe2018overview}. Such missions adopted warm-gas propulsion systems or cold-gas multi-thrusters to perform RPO for CubeSats.

The EP-Gemini mission is designed to push this boundary by demonstrating CubeSat RPO with a single resistojet as the propulsion unit.
The EP-Gemini mission is an academic cooperation project to advance Taiwanese space industry and satellite design capability. This mission, consisting of two 3U CubeSats, will be the first RPO mission in Taiwanese space history.
Each satellite bus includes commercial off-the-shelf parts, as well as custom parts from local suppliers. This mission is scheduled to launch to LEO in 2024. 

In this work, we introduce the control framework of the EP-Gemini mission. In particular, a CubeSat chaser shall be controlled to establish a safety ellipse relative to the other uncooperative CubeSat target. Although several methods, see References~\citenum{gaylor2007algorithms, navabi2013algebraic, shuster2020analytic, garcia2021electric}, have been proposed for similar passively safe RPO, there exist special considerations pertinent to the EP-Gemini mission. Such relevant factors are discussed and incorporated into the EP-Gemini control problem. To solve this EP-Gemini problem, we first examine the relationship between safety ellipse and relative mean orbit elements. After defining the desired relative mean orbit elements, we design four basic maneuver blocks that achieve two goals simultaneously: 1) tracking a relative mean orbit element independently, and 2) ensuring that the constraints of the EP-Gemini problem are satisfied. Lastly, the control framework is constructed by applying these maneuver blocks in a specific sequence to guarantee safe RPO. The contributions of this work include: 1) package the RPO maneuver into basic maneuver blocks considering the operation of electric propulsion units, and 2) demonstrate the feasibility of a safety ellipse mission for two CubeSats with indirect information link.

We implement this control framework in the environment of the Ansys Systems Tool Kit (STK) and demonstrate its efficacy for the EP-Gemini mission. We make some remarks on the results of the numerical simulation, and discuss the limitations and future improvement of the proposed control framework.

\section{Background}
The EP-Gemini mission will demonstrate CubeSat rendezvous and proximity operations (RPO) with an uncooperative target using electric propulsion in low Earth orbit (LEO). The mission consists of two 3U CubeSats: 1) a target that has no propulsion unit, and 2) a chaser that has a single electric propulsion thruster. Initially, the target and the chaser are launched to the same orbit as a single object. Upon the separation time, the two satellites separate from the upper stage of the rocket, enter different orbits, and begin their commissioning. After the commissioning is complete, the chaser performs a passively safe RPO to establish a safety ellipse relative to the target. The chaser circumnavigates the target until performing a final maneuver to exit the vicinity of the target.

The control framework of the EP-Gemini mission is driven by the separation mechanism and the system design of the chaser satellite. Firstly, the two satellites will be separated by the Poly Picosatellite Orbital Deployer (P-POD) with an estimated 2 m/s relative separation speed. This separation mechanism, along with the estimated 30-days commissioning period, determines the possible initial conditions of the RPO task. Secondly, the chaser is designed to have an initial wet mass of 4 kg and operate on a body-fixed 20 watt resistojet. This resistojet consumes electricity and generates a maximum 6 micronewtons thrust at a specific impulse of 100 seconds; it provides a total specific impulse of 270 seconds, i.e., $\Delta V_{\max}=67.5$ m/s. The Attitude Determination and Control System (ADCS) of the chaser is capable of 3-axis stabilization. The electric power system of the chaser includes a 45 watt-hours battery and two solar array wings with no gimbals. Lastly, the two satellites will downlink their position and velocity data in the Earth-centered inertial coordinate system. 

\subsection{Safety Ellipse}
Let $\vec{r}_1$ represent the position vector of the target described in the Earth-centered inertial frame (ECI). Then the RSW satellite coordinate system of the target, denoted as $[\hat{R}_1,\hat{S}_1,\hat{W}_1]$, is centered at the target position, and the unit vectors $\hat{R}_1,\hat{S}_1,\hat{W}_1$ are:

\begin{equation} \label{eq:RSW}
    \begin{aligned}
    \hat{R}_1 &= \frac{\vec{r}_1}{||\vec{r}_1||_2} \\
    \hat{S}_1 &= \hat{W}_1 \times \hat{R}_1\\
    \hat{W}_1 &= \frac{\vec{h}_1}{||\vec{h}_1||_2},\; \vec{h}_1 = \vec{r}_1 \times \dot{\vec{r}}_1.
    \end{aligned}
\end{equation}

Let $\vec{r}_2$ represent the position vector of the chaser described in the ECI. Then the motion of the chaser relative to the target, $\vec{r}(t)=\vec{r}_2(t)-\vec{r}_1(t)=x(t)\hat{R}_1+y(t)\hat{S}_1+z(t)\hat{W}_1$, is approximated by the Hill's equations under certain assumptions \cite{vallado2013fundamentals}:
\begin{equation} \label{eq:hills}
    \begin{aligned}
    \ddot{x}-2n\dot{y}-3n^2x &= u_x \\
    \ddot{y}{+}2n\dot{x} &= u_y \\
    \ddot{z}+n^2z &= u_z,
    \end{aligned}
\end{equation}
where $n$ is the mean motion of the target, $x$ is the relative radial distance, $y$ is the relative along-track distance, $z$ is the relative cross-track distance, and $u_x,u_y,u_z$ are the thrust accelerations along the $\hat{R}_1,\hat{S}_1,\hat{W}_1$ axes. The analytical solution exists for the unforced case $(u_x=u_y=u_z=0)$, and a static ellipse that centers at the origin of the $[\hat{R}_1,\hat{S}_1,\hat{W}_1]$ is formed at $t_0$ if:
\begin{equation} \label{eq:safety_ellipse}
    \begin{aligned}
    \dot{x}(t_0) &= \frac{n}{2}y(t_0) \\
    \dot{y}(t_0) &= -2nx(t_0) \\
    \end{aligned}
\end{equation}

\begin{figure}[htb]
\centering
\includegraphics[width=.7\textwidth]{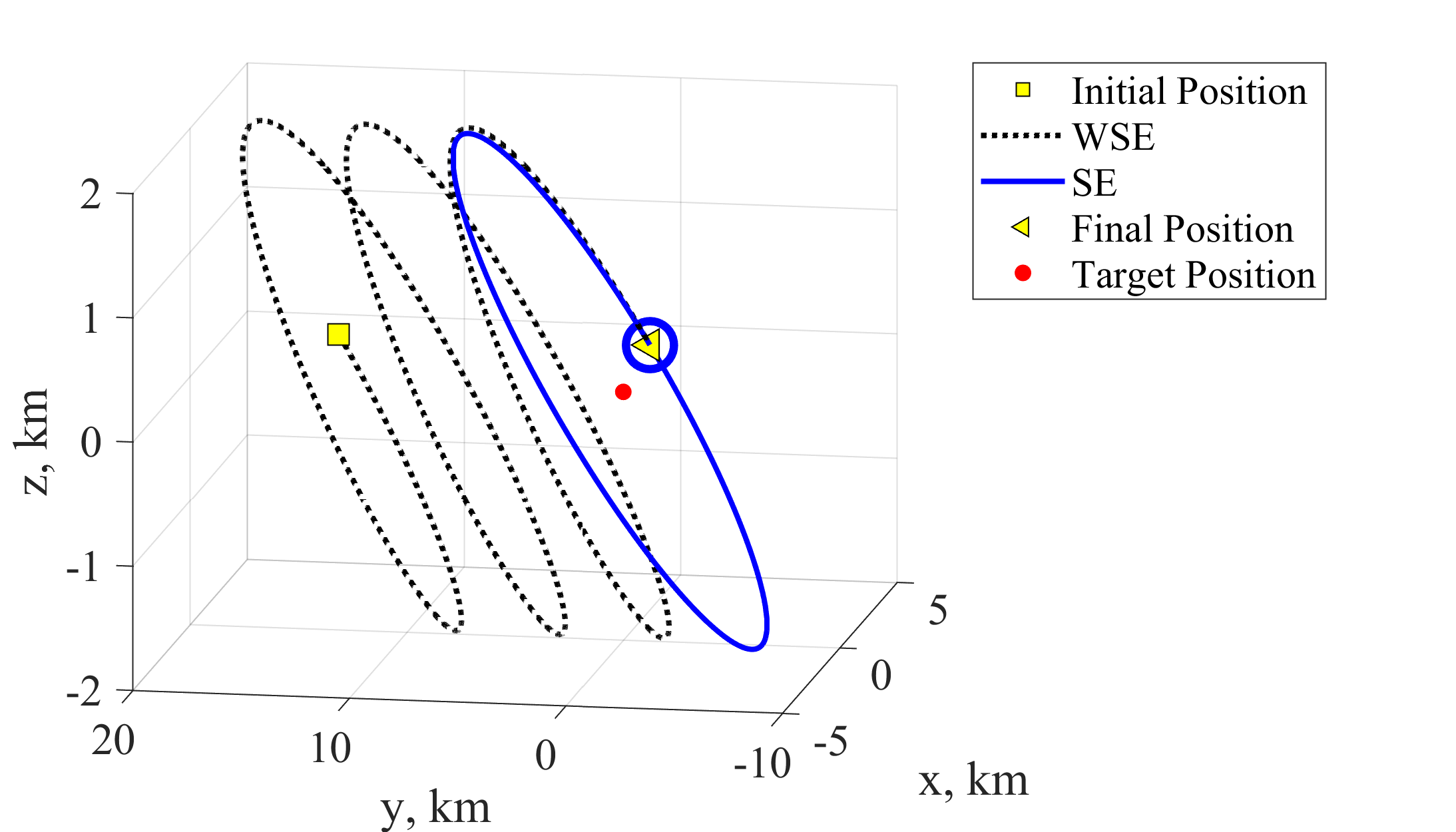} 
\caption{Exemplary safety ellipse maneuver}
\label{fig:WSE_SE}
\end{figure}

Based on the static ellipse (SE), a safety ellipse is established by phasing the $z$ motion (a harmonic motion) and the $x$-$y$ motion in Eq.~\eqref{eq:hills} such that 1) the chaser intersects the $xy$-plane near the ($x=x_{\max},y=0$) points, and 2) the ellipse is out-of-plane from the $xy$-plane. Another relative motion, called walking safety ellipse (WSE), drifts the center of the safety ellipse towards the target. Figure~\ref{fig:WSE_SE} illustrates an exemplary safety ellipse maneuver propagated by Eq.~\eqref{eq:hills}. The result shows that the chaser initially drifts towards the origin over 3 periods using the WSE. Then it maintains the static safety ellipse over 10 periods, where the blue hollow dot and yellow triangle indicate the start and end of such a circumnavigation, respectively.

\section{Methods}

\subsection{EP-Gemini Problem Formulation}
A general safety ellipse problem is: given an initial relative state of the chaser $s_i=[x_i,y_i,z_i,\dot{x}_i,\allowbreak \dot{y}_i,\dot{z}_i]^T$, find a control sequence $\vec{u}(t), t\in[t_i,t_f]$, such that the state trajectory over the time interval between the initial and final states ensures a low collision risk (even if thrust authority is lost), and the final state $s_f$ establishes a static safety ellipse with $\vec{u}(t) \rightarrow 0$ as $t \rightarrow t_f$. By introducing an objective function, this problem can be formulated as an optimization problem, and several optimal feedback laws have been proposed.\cite{garcia2021electric,bando2013plane,willis2017relative} However, it is unlikely to directly apply existing optimal solutions to the EP-Gemini mission due to its concept of operation and system constraints. The special considerations of our mission are:
\begin{enumerate}
    \item \textbf{Body-fixed thrust vector:}
    Since the chaser uses a body-fixed electric propulsion thruster, changing the thrust direction requires changing the satellite's attitude. Thus, it is difficult to realize a control sequence that varies the thrust vector direction frequently. Let $[\hat{R}_2,\hat{S}_2,\hat{W}_2]$ be the RSW coordinate system of the chaser. We consider a practical EP-Gemini RPO control consisting of piecewise constant thrust commands $\vec{u}(t_j)$ with constant thrust along one of the unit vectors of the $\hat{R}_2,\hat{S}_2,\hat{W}_2$ over the j-th time interval:
    \begin{equation}
    \vec{u}(t_j) =
        \begin{cases}
        F_{R}\hat{R}_2 + 0\hat{S}_2 + 0\hat{W}_2\\
        0\hat{R}_2 + F_{A}\hat{S}_2 + 0\hat{W}_2\\
        0\hat{R}_2 + 0\hat{S}_2 + F_{C}\hat{W}_2,\\
        \end{cases},\; t_j\in[t_{j,i},t_{j,f}],\; t_i\leq t_{j,i} \leq t_{j,f} \leq t_f      
    \end{equation}
    \item \textbf{Power limitation of electric propulsion:} 
    When the electric propulsion thruster is firing, it is not possible to simultaneously direct the body-fixed solar array towards the sun. Accordingly, the continuous firing time of the electric propulsion is constrained by the battery capacity. Given the electric propulsion unit, the estimated power consumption of other subsystems, and the estimated end-of-life battery capacity, the maximum time allowed for one continuous firing is 15 minutes. This constraint leads to a practical maneuver with a series of thruster firings separated by battery charging phases, which is illustrated in Figure~\ref{fig:firing_constraint}.

\begin{figure}[htb]
\centering
\begin{subfigure}{0.8\textwidth}
\includegraphics[width=1.0\linewidth]{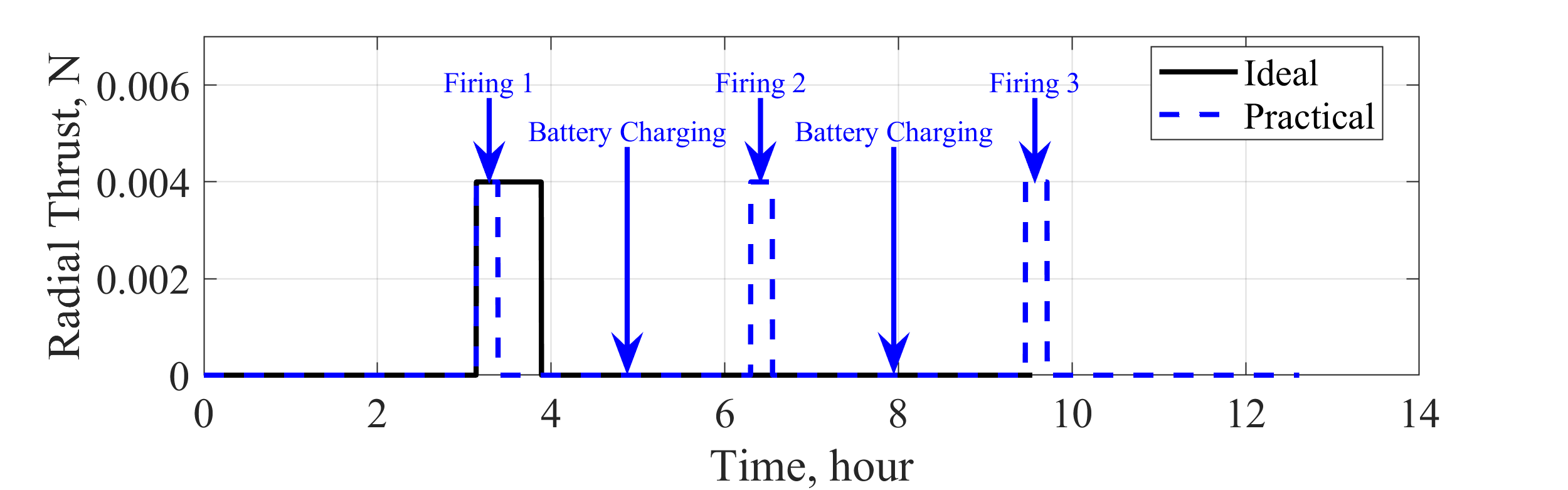}
\caption{Control thrust over time}
\end{subfigure}
\vfill
\begin{subfigure}{0.8\textwidth}
\includegraphics[width=1.0\linewidth]{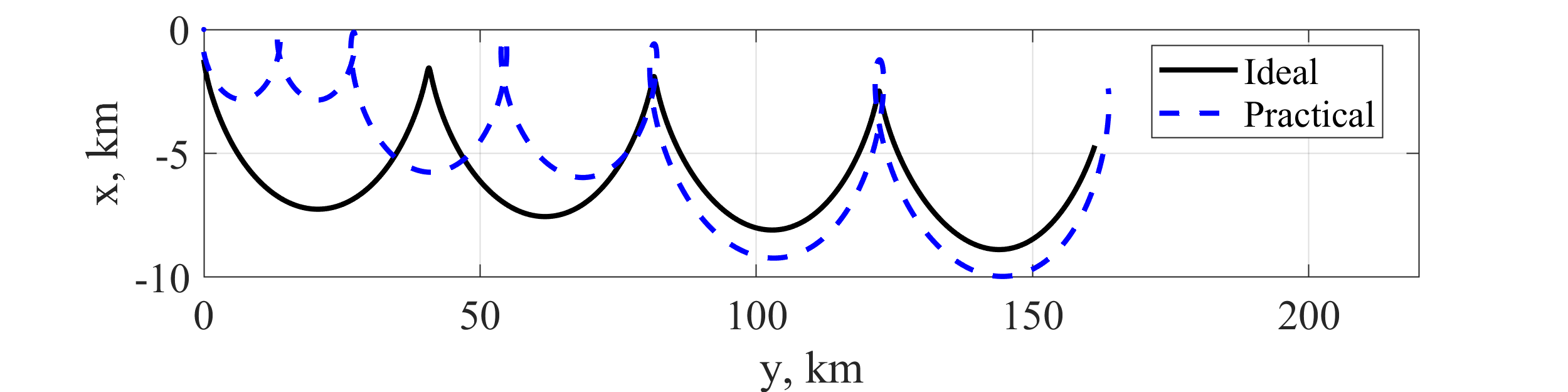}
\caption{Relative trajectory}
\end{subfigure}
\caption{Comparison between unconstrained (ideal) and constrained (practical) thrust control.}
\label{fig:firing_constraint}
\end{figure}

    \item \textbf{Temporal limitation of orbit determination:} The 3U volume dimension restricts the usage of relative position instrument, e.g., inter-satellite link or optical range sensor. As a result, we consider a practical control framework that takes the mean orbit elements as inputs. Using mean orbit elements permits stable tracking because the short period oscillation is averaged out compared to osculating orbit elements. Such a property of mean orbit elements is leveraged to address the temporal limitation of state update. In practice, the concept of the data flow is: First, the target satellite will downlink the Global Navigation Satellite Systems (GNSS) data to the ground station. Secondly, this data will be used to calculate osculating orbit elements, which is then transformed into mean orbit elements. Thirdly, the processed mean orbit elements of the target will be uplinked to the chaser to inform the state of the target. Lastly, the onboard processor of the chaser utilizes this information to control its RPO maneuver. 
    Due to the processes of downlink, orbit element transformation, and uplink, the average update rate of the relative orbit information is estimated to be 175 minutes.
\end{enumerate}
The EP-Gemini problem therein is reformulated as a typical safety ellipse problem subject to the above constraints on the control inputs and the temporal limitation of orbit determination. Note that the optimization problem of the EP-Gemini is left for future work.

\subsection{Desired Orbit Formulation}
\begin{figure}[htb]
\centering
\begin{subfigure}{0.45\textwidth}
\includegraphics[width=1.0\linewidth]{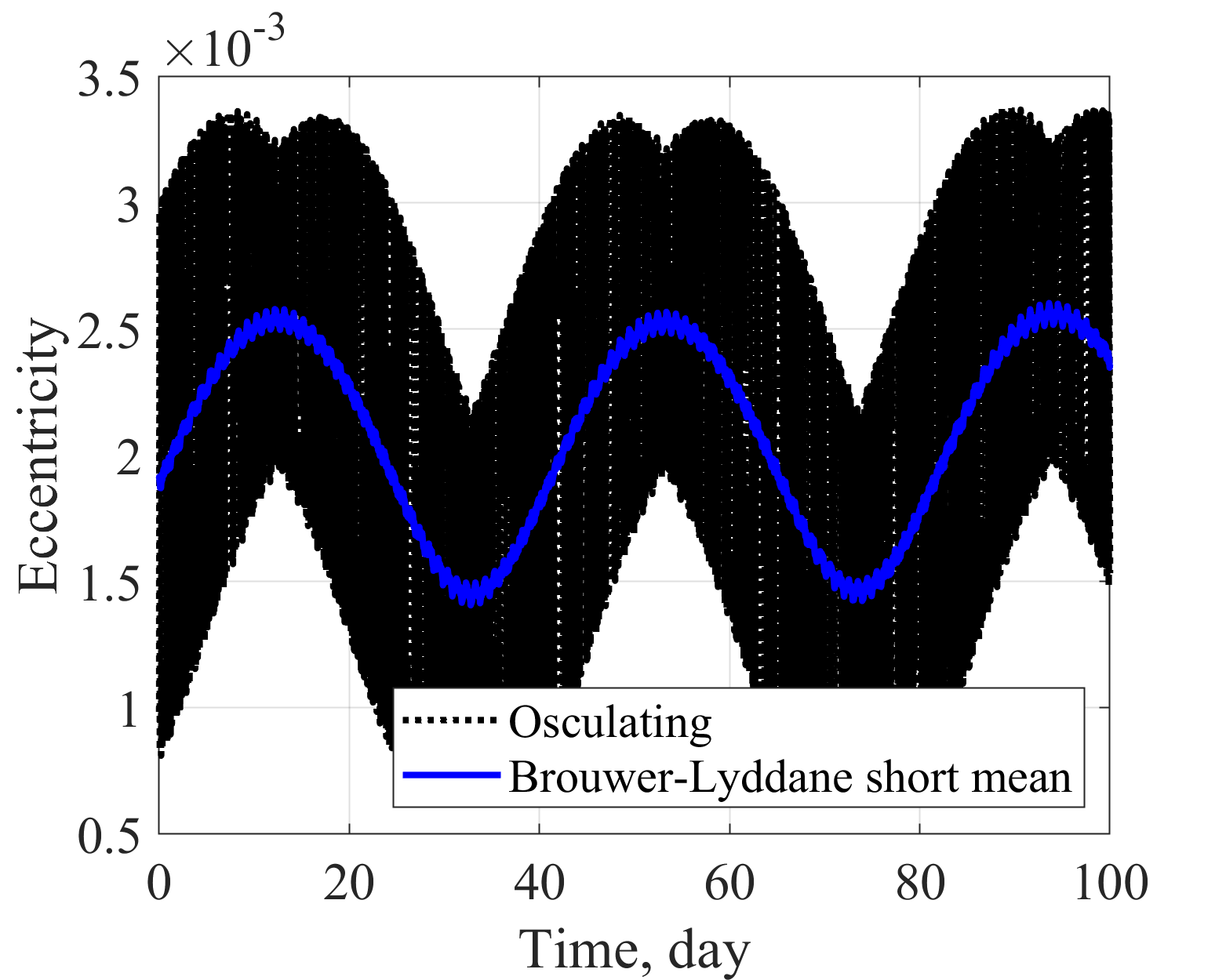}
\caption{Eccentricity over time}
\label{fig:OE_e}
\end{subfigure}
\hspace{0cm}
\begin{subfigure}{0.45\textwidth}
\includegraphics[width=1.0\linewidth]{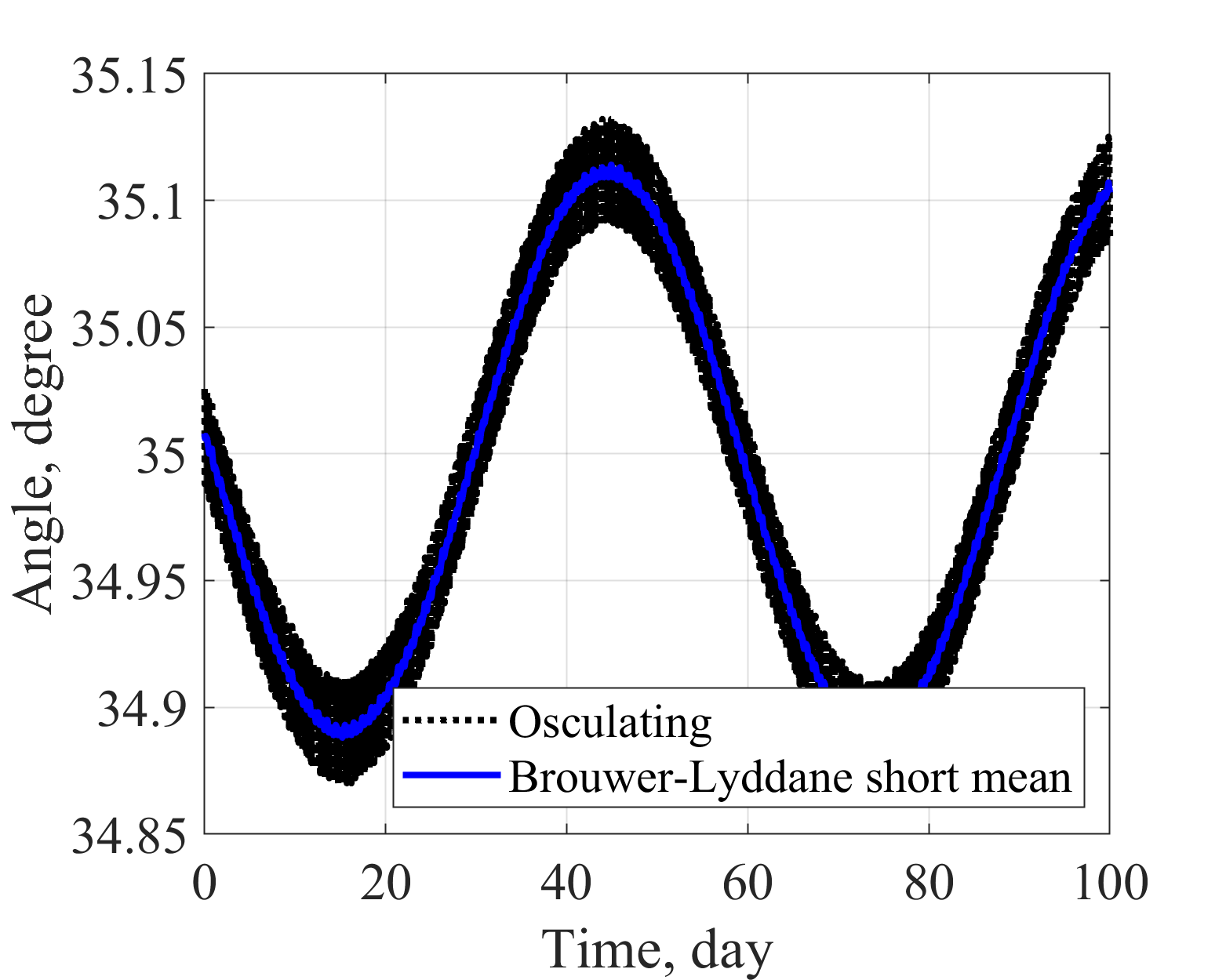}
\caption{Inclination over time}
\label{fig:OE_i}
\end{subfigure}
\caption{Simulated osculating and mean orbit elements of the target satellite}
\label{fig:OE_suggestion}
\end{figure}

\begin{figure}[htb]
\centering
\begin{subfigure}{0.45\textwidth}
\includegraphics[width=1.0\linewidth]{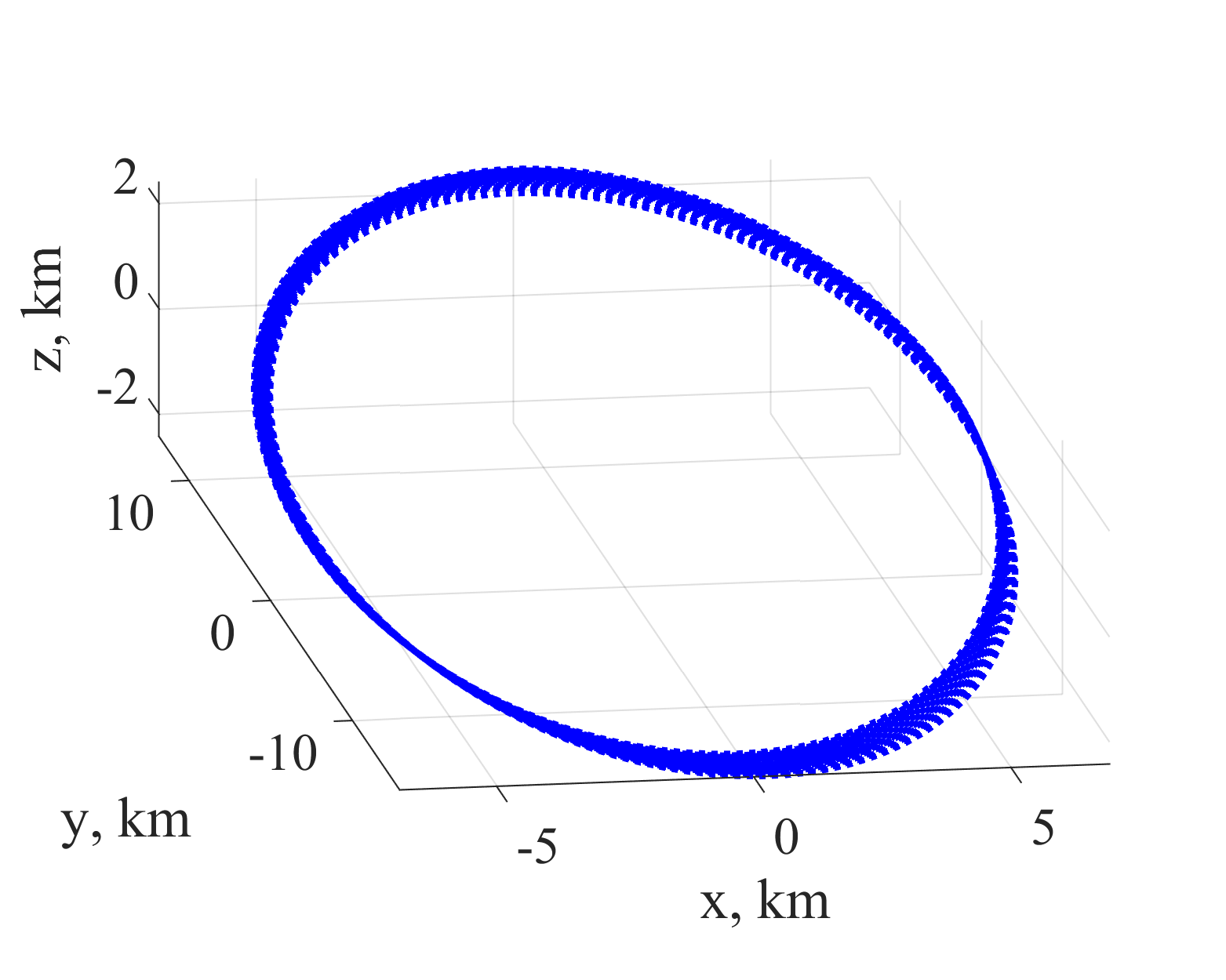}
\caption{3D relative trajectory}
\end{subfigure}
\hspace{0cm}
\begin{subfigure}{0.45\textwidth}
\includegraphics[width=1.0\linewidth]{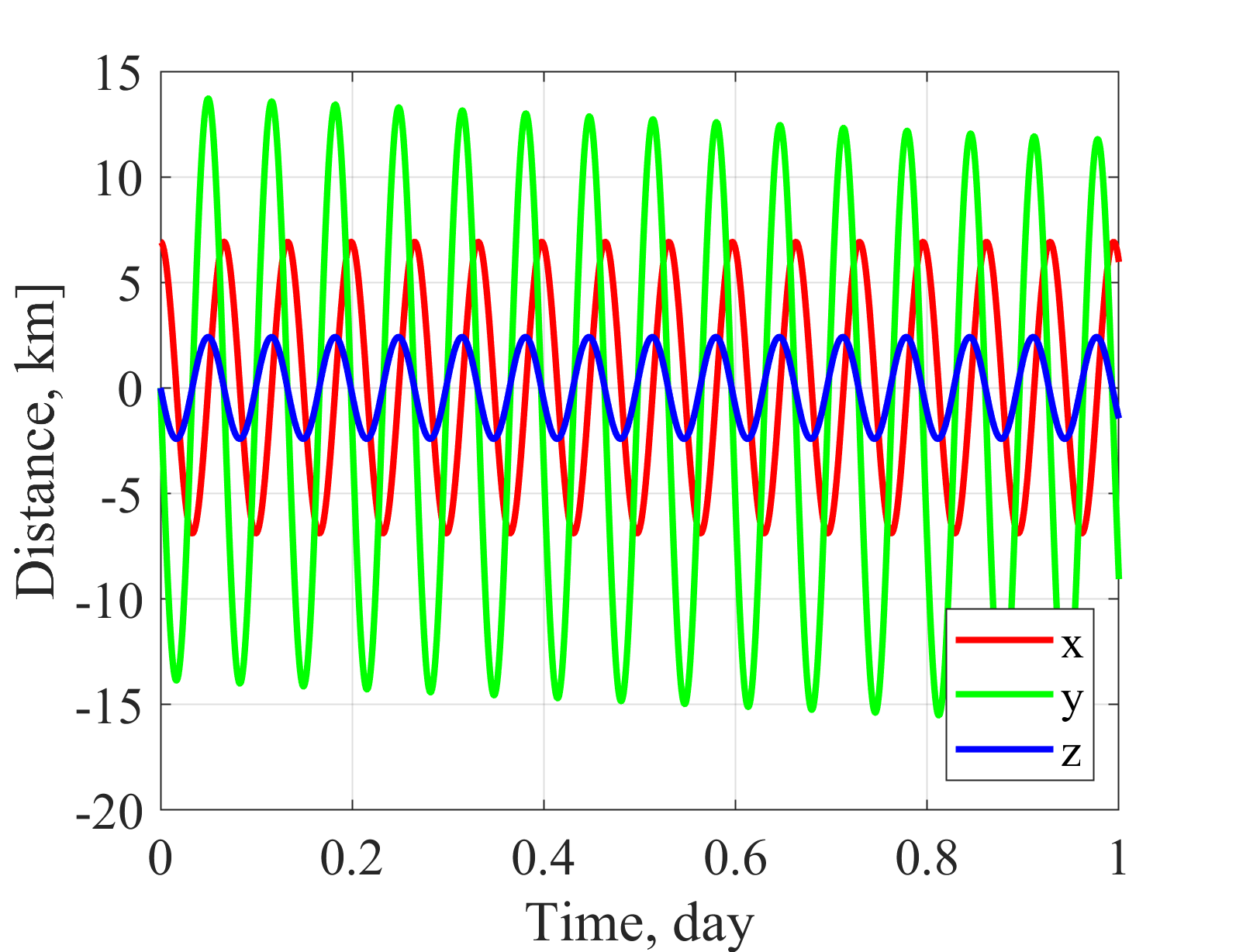}
\caption{Relative distance over time}
\end{subfigure}
\caption{Ideal desired safety ellipse}
\label{fig:ideal_SE}
\end{figure}

The desired safety ellipse can be established by converging the chaser's mean orbit elements to desired values.\cite{schaub2004relative,schaub2000spacecraft,6581216} The chaser's desired mean orbit elements are obtained by:
\begin{equation}
	\label{eq:ab}
	\text{OE}_{2,\text{des}}= \text{OE}_1 + \Delta \text{OE}_{\text{des}},
\end{equation}
where $\text{OE}=\{a,e,i,\Omega,u\}$ denotes the mean orbit elements, $\text{OE}_{\text{2,des}}$ is the chaser's desired mean orbit elements, $\text{OE}_1$ is the target's mean orbit elements, and $\Delta \text{OE}_{\text{des}}$ is the relative desired mean orbit elements (note that $\Delta \text{OE} = \text{OE}_2 - \text{OE}_1$ is the relative actual mean orbit elements). In this work, we require $\Delta a_{\text{des}} = \Delta \Omega_{\text{des}} = \Delta u_{\text{des}} = 0$, where $\Delta u_{\text{des}} = \Delta \theta_{\text{des}} + \Delta \omega_{\text{des}}$ is the relative desired mean argument of latitude. For the relative eccentricity and inclination, the desired values ($\Delta e_{\text{des}}$ and $\Delta i_{\text{des}}$) are determined by simulating the target orbit. Figure~\ref{fig:OE_suggestion} shows the osculating and mean orbit elements of the target satellite in the simulation environment of Ansys Systems Tool Kit (STK). By examining the difference between the mean and osculating orbit elements in Figure~\ref{fig:OE_suggestion}, we suggest the minimum $\Delta e_{\text{des}} = 0.001$ and $\Delta i_{\text{des}} = 0.02$ degree as a baseline for the EP-Gemini mission. 
After defining the $\Delta \text{OE}_{\text{des}}$, an ideal desired safety ellipse of the EP-Gemini can be mapped from the $\text{OE}_1$. Figure~\ref{fig:ideal_SE} illustrates the desired safety ellipse with the dimension of 14 km $\times$ 28 km in the x-y plane, and an amplitude of 5 km in the z direction.

In addition, formulating the desired safety ellipse in terms of mean orbit elements is the key to solving the temporal limitation mentioned in the EP-Gemini problem. Within the average state update rate of 175 minutes, the mean orbit elements have relatively small variation; for example, see the oscillation period of mean eccentricity and mean inclination in Figure~\ref{fig:OE_suggestion}. This property allows us to set the chaser's desired state as constant until the next orbit information is obtained.

\newpage
\subsection{Maneuver Blocks}
Considering the control limitations of the EP-Gemini problem, we design four basic maneuver blocks to correct the chaser's mean orbit elements to the desired values: $\text{OE}_2 \rightarrow \text{OE}_{2,\text{des}}$ (or $\Delta \text{OE} \rightarrow \Delta \text{OE}_{\text{des}}$). In these maneuver blocks, the time intervals of all the thruster firings are shorter than 15 minutes (i.e., the maximum thruster firing time), and at least a battery charging phase (over a full orbit cycle) is applied between two successive thruster firings. The four maneuver blocks and their thruster firings described in the chaser coordinate $[\hat{R}_2,\hat{S}_2,\hat{W}_2]$ are:
\begin{enumerate}
    \item \textbf{RAAN correction maneuver ($\Omega_{\text{cor}}$)}: 
    In this maneuver block, the chaser does an altitude maneuver and leverages the J2 perturbation to correct the right ascension of the ascending node (RAAN) difference accumulated over the commissioning phase prior to the RPO. This maneuver consists of a series of along-track firings, an unforced propagation, and a series of reversed along-track firings. The first series of along-track firings are used to initiate opposite RAAN drift with respect to the prior commissioning phase. The accumulated RAAN difference is then reduced during the unforced propagation. Lastly, a series of reversed along-track firings with respect to the first firings are used to compensate the altitude difference between the chaser and the target. The direction of the first series of along-track firings depends on the initial condition of the RPO. For example, if the initial condition of the RPO is such that the chaser's altitude is greater than the target, then the first series of along-track firings will be positive, and vice versa.

    \item \textbf{Argument of latitude correction maneuver ($u_{\text{cor}}$)}:
    In this maneuver block, the chaser changes its altitude and leverages orbit period difference to reduce the relative along-track distance. This maneuver block consists of a series of along-track firings, a short period of unforced propagation, and a series of along-track firings in the reversed direction. Compared to the $\Omega_{\text{cor}}$ maneuver block, the unforced propagation time in the $u_{\text{cor}}$ maneuver block is relatively short in order to minimize the change of chaser's RAAN due to the J2 perturbation.

    \item \textbf{Inclination correction maneuver ($i_{\text{cor}}$)}:
    This maneuver block consists of a series of cross-track firings upon flying by the ascending or descending node, which depends on the relative desired inclination. Without loss of generosity, the chaser will perform positive/negative cross-track firings after the ascending node to increase/decrease its inclination. Figure~\ref{fig:inc_block} shows the result of applying a $i_{\text{cor}}$ maneuver block. 
    \begin{figure}[htb]
    \centering
    \begin{subfigure}{0.8\textwidth}
    \includegraphics[width=1.0\linewidth]{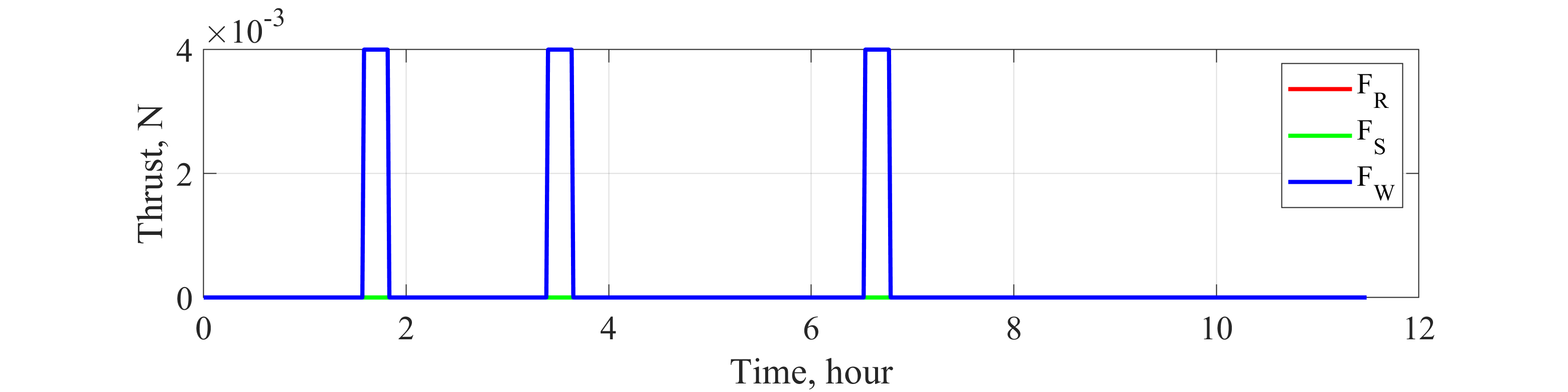}
    \caption{Control thrust over time}
    \end{subfigure}
    \vfill
    \begin{subfigure}{0.8\textwidth}
    \includegraphics[width=1.0\linewidth]{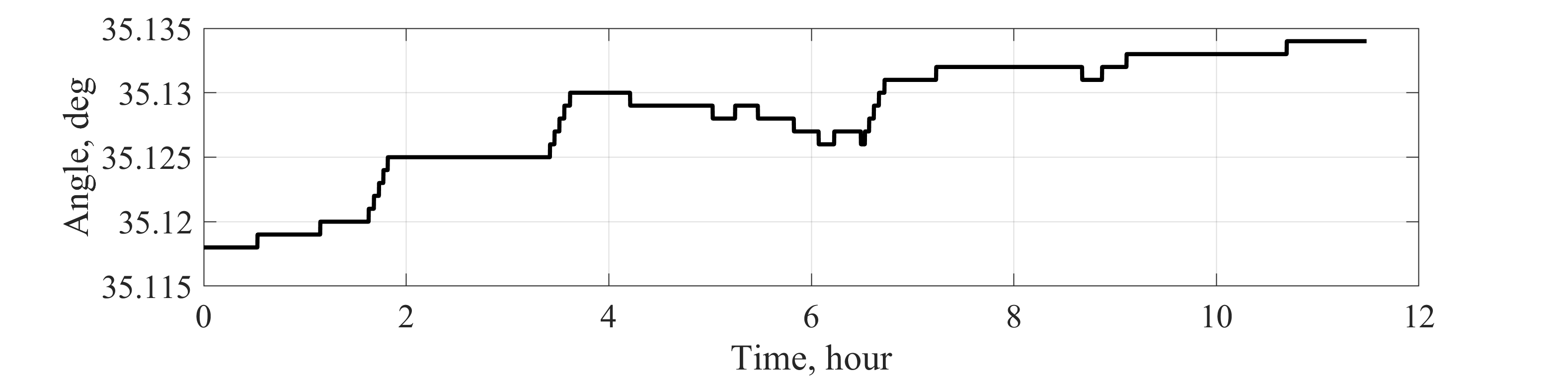}
    \caption{Inclination over time}
    \end{subfigure}
    \begin{subfigure}{0.8\textwidth}
    \includegraphics[width=1.0\linewidth]{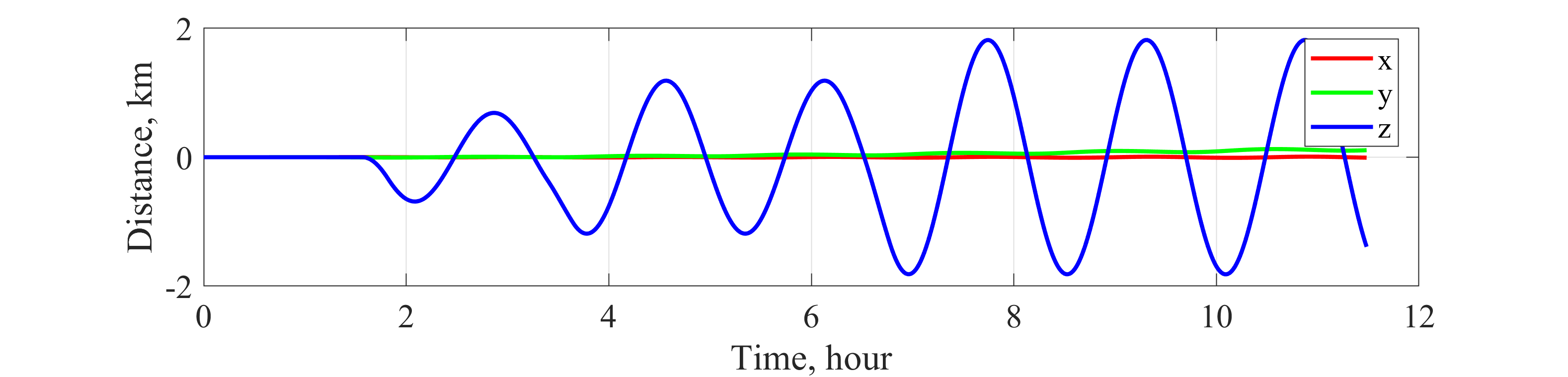}
    \caption{Relative distance over time}
    \end{subfigure}
    \caption{Control and effects of the inclination correction maneuver block}
    \label{fig:inc_block}
    \end{figure}

    \item \textbf{Eccentricity correction maneuver ($e_{\text{cor}}$)}:
    This maneuver block includes two operations in sequence. The first operation consists of three firings: a positive along-track firing at the perigee over a time interval $t_{e1}$, and two negative along-track firings at the $\frac{1}{4}$ and the $\frac{3}{4}$ orbit locations, each fires over $\frac{1}{2}t_{e1}$ time interval. The negative firings are used to compensate the altitude change caused by the first positive along-track firing without affecting the eccentricity. Following the first operation, the second operation consists of a negative along-track firing at the apogee over a time interval $t_{e2}$, and two positive along-track firings at the $\frac{1}{4}$ and the $\frac{3}{4}$ orbit location, each fires over $\frac{1}{2}t_{e2}$ time interval. These two operations allows the chaser to correct its eccentricity while in the meantime exhibit small along-track offset. Figure~\ref{fig:ecc_block} shows the result of applying one $e_{\text{cor}}$ maneuver block.
    \begin{figure}[htb]
    \centering
    \begin{subfigure}{0.8\textwidth}
    \includegraphics[width=1.0\linewidth]{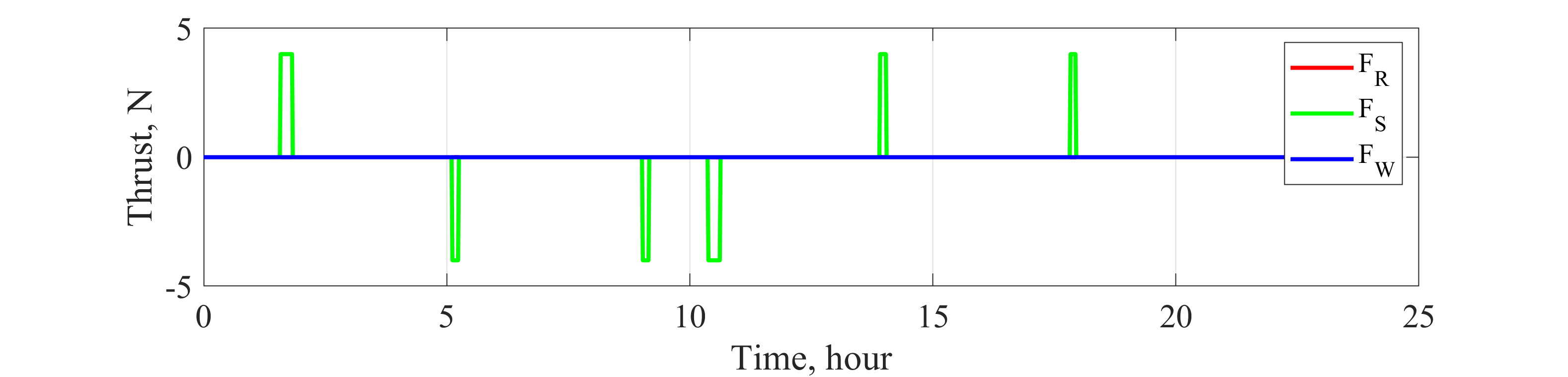}
    \caption{Control thrust over time}
    \end{subfigure}
    \vfill
    \begin{subfigure}{0.8\textwidth}
    \includegraphics[width=1.0\linewidth]{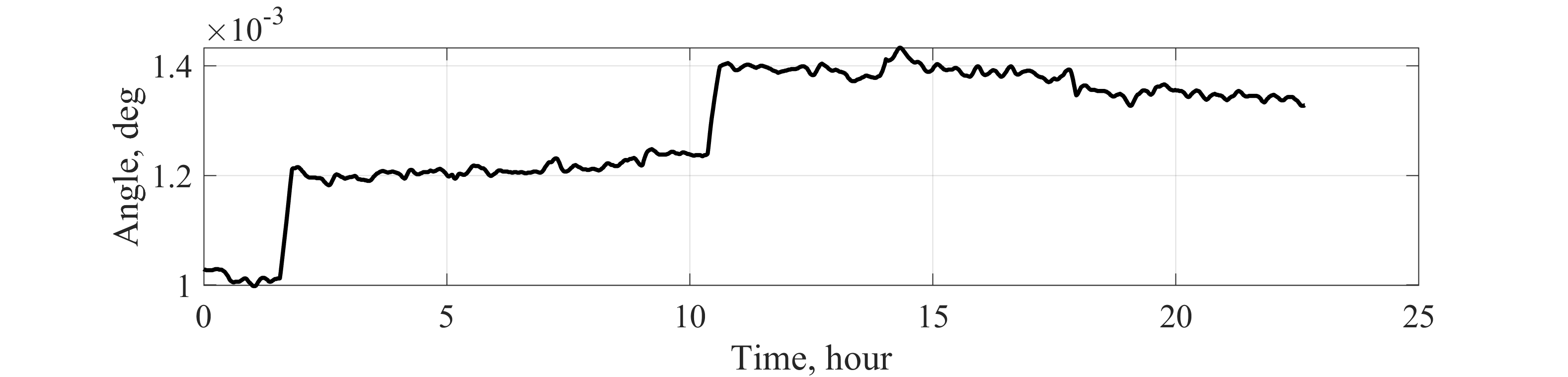}
    \caption{Eccentricity over time}
    \end{subfigure}
    \begin{subfigure}{0.8\textwidth}
    \includegraphics[width=1.0\linewidth]{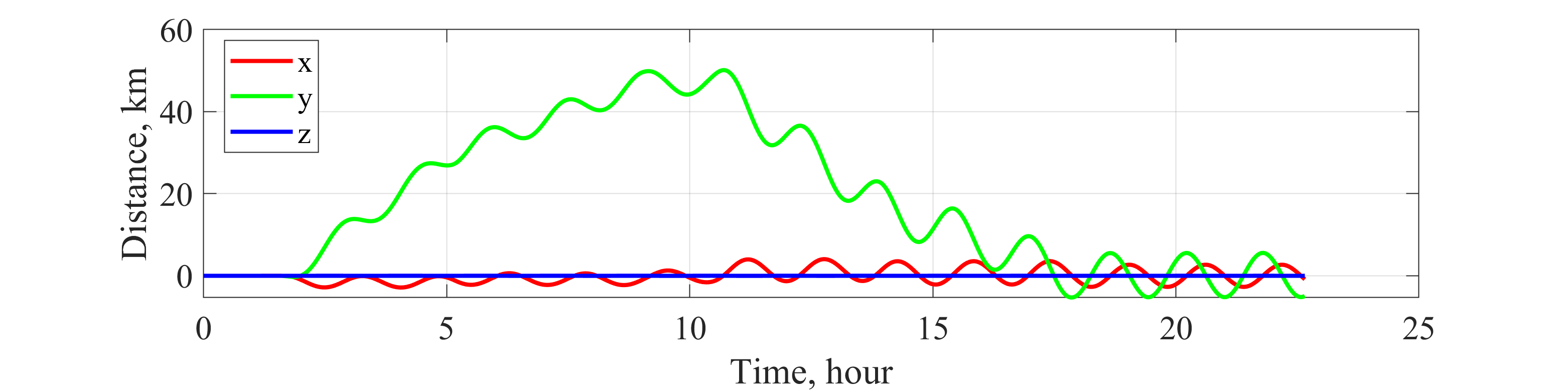}
    \caption{Relative distance over time}
    \end{subfigure}
    \begin{subfigure}{0.8\textwidth}
    \includegraphics[width=1.0\linewidth]{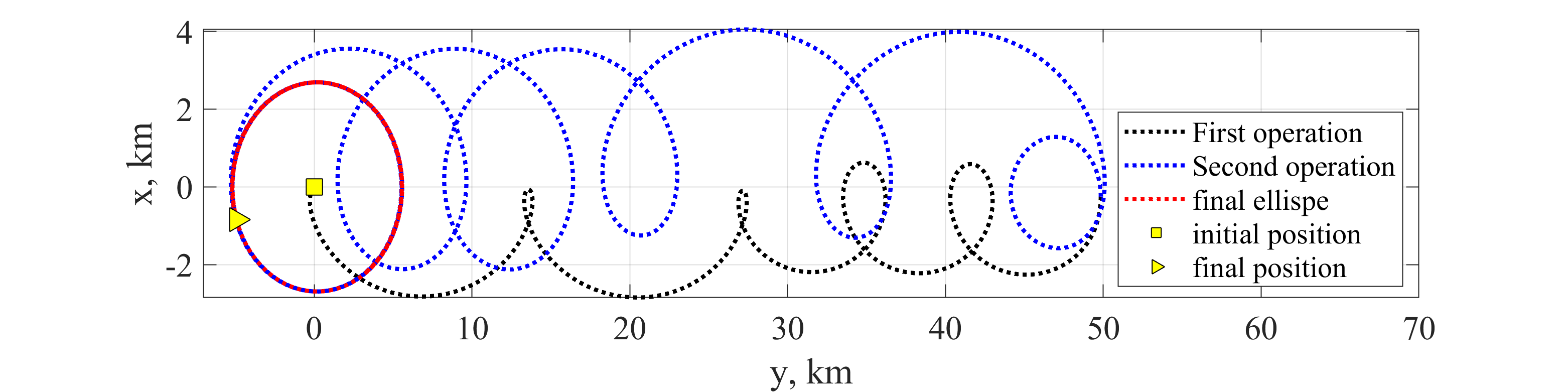}
    \caption{Relative trajectory of the chaser on cross-track and radial plane}
    \end{subfigure}
    \caption{Control and effects of the eccentricity correction maneuver block}
    \label{fig:ecc_block}
    \end{figure}
    
\end{enumerate}

\newpage
\subsection{EP-Gemini Control Framework}
\begin{figure}[htb]
\centering
\includegraphics[width=1.0\textwidth]{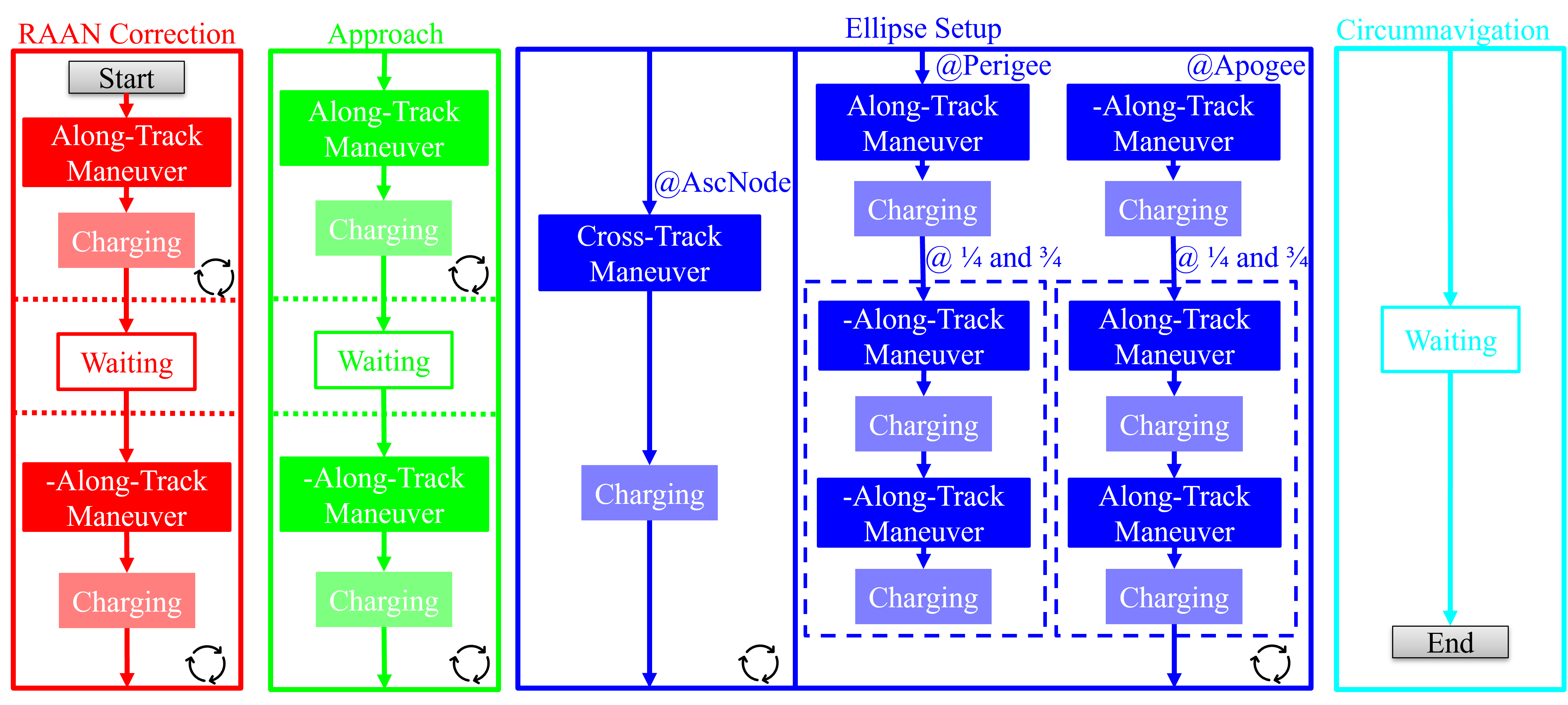} 
\caption{EP-Gemini control framework}
\label{fig:RPO_seq}
\end{figure}

The EP-Gemini control framework includes four subsequent phases illustrated in Figure~\ref{fig:RPO_seq}. In each phase, the basic maneuver blocks are applied recursively such that certain relative states converge to the desired values.
\begin{enumerate}
    \item \textbf{RAAN Phase:}
    This phase begins after the commissioning is complete. In this phase, the chaser applies the $\Omega_{\text{cor}}$ maneuver block to correct its RAAN until $\Delta \Omega \rightarrow \Delta \Omega _{\text{des}} = 0$ degree. Note that the implementation of $\Omega_{\text{cor}}$ maneuver block ensures that $\Delta a \rightarrow \Delta a_{\text{des}} = 0$ at the end of this phase.

    \item \textbf{Approach Phase}: 
    Following the end of the RAAN phase, the chaser applies the $u_{\text{cor}}$ maneuver block to approach the target until the relative along-track distance is less than 50 km. This reserved along-track distance is critical to the safety of the EP-Gemini RPO because the out-of-plane separation between the chaser and the target has not been established. Upon the completion of this phase, the relative along-track drift is minimized by implementing the $u_{\text{cor}}$ maneuver block, in which $\Delta a \rightarrow \Delta a_{\text{des}} = 0$.

    \item \textbf{Ellipse Setup Phase:}
    After the end of the approach phase, the chaser applies the $i_{\text{cor}}$ and $e_{\text{cor}}$ maneuver blocks to establish the desired safety ellipse. The $i_{\text{cor}}$ maneuver block is first applied recursively until $\Delta i \rightarrow \Delta i_{\text{des}}=0.02$ degree, which creates the relative out-of-plane motion. Next, the $e_{\text{cor}}$ maneuver block is applied recursively until $\Delta e \rightarrow \Delta e_{\text{des}}=0.001$, creating the relative ellipsoidal motion in the x-y plane. Additionally, the reserved along-track distance is compensated in this phase by leveraging the along-track offset of the first or second operation in the $e_{\text{cor}}$ maneuver block. For example, applying the first operation without the second operation induces positive along-track offset, and vice versa. This phase ends when the desired safety ellipse is established (with the center of the ellipse approaching the target).

    \item \textbf{Circumnavigation Phase:}
    The chaser begins the circumnavigation phase after the end of the ellipse setup phase. In this phase, we assume no thruster firings.
\end{enumerate}

\section{Numerical Simulation}
We demonstrate the efficacy of our control framework in the environment of Ansys Systems Tool Kit (STK), where high order perturbations, such as J2 perturbation, atmospheric drag, and solar radiation pressure, are simulated. The initial orbit elements for the target is: ($a,e,i,\Omega,\omega,\theta$)=($6925.68\allowbreak \text{ km},0.0019\text{ deg},35.008\text{ deg},3.006\text{ deg},0\text{ deg},0\text{ deg}$). At $t=0$, the separation occurs, which induces a change of 2.65 km semi-major axis and 0.00043 eccentricity in the chaser orbit. After 30 days of commissioning, the chaser applies the control framework to perform the rendezvous and proximity operation (RPO). The simulation terminates when the circumnavigation phase exceeds 30 days, which allows us to examine the time evolution of the safety ellipse over the nominal circumnavigation (3 days) and the extended circumnavigation (30 days).

\begin{figure}[htb]
\centering
\includegraphics[width=1.0\textwidth]{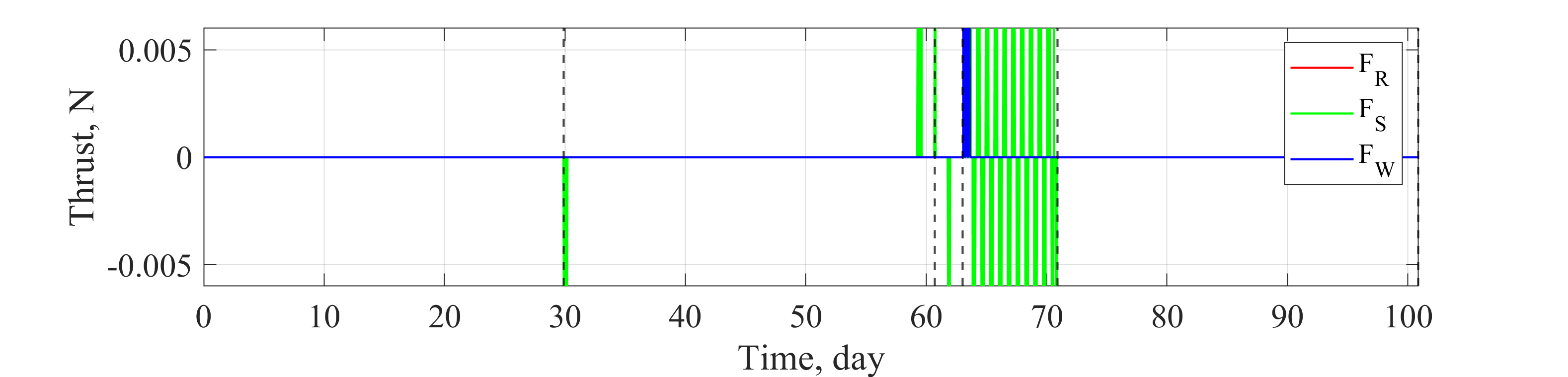} 
\caption{Control thrust over time}
\label{fig:final_Thrust}
\end{figure}

\begin{figure}[htb]
\centering
\includegraphics[width=1.0\textwidth]{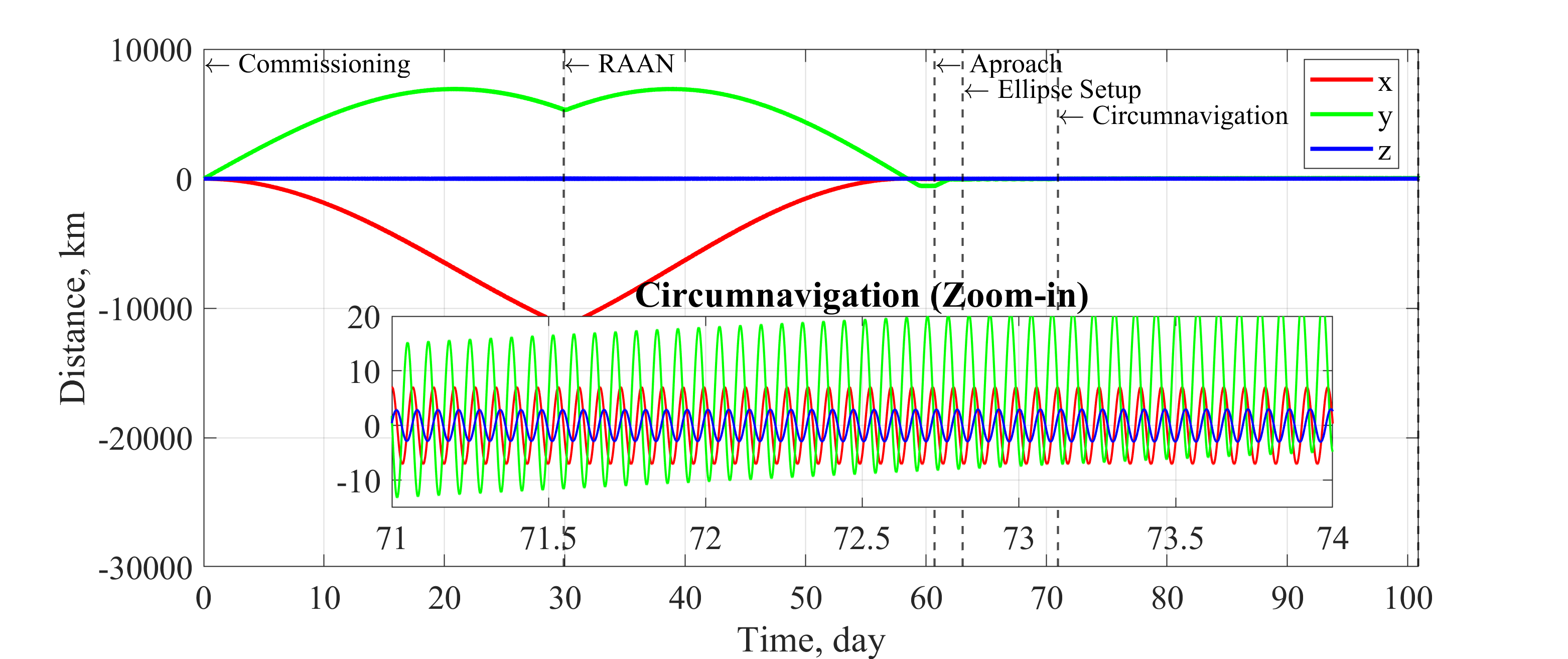} 
\caption{Relative distance over time}
\label{fig:final_RSW}
\end{figure}

Figure~\ref{fig:final_Thrust} and~\ref{fig:final_RSW} illustrate the chaser's control thrust and the resulting relative distance. Note that according to the EP-Gemini control framework, no radial firing $F_R$ exists in Figure~\ref{fig:final_Thrust}. In Figure~\ref{fig:final_RSW}, the $x(t)$ and $y(t)$ increase during the commissioning. These accumulated difference are both reduced from the order of thousands of kilometer to hundreds of kilometer during the first RAAN phase of the RPO. The zoom-in plot in Figure~\ref{fig:final_RSW} shows the relative distance during the initial 3 days of the circumnavigation phase; the oscillation amplitude in the radial, along-track, and cross-track directions are about 14 km, 27 km, and 8 km, respectively.

\begin{figure}[htb]
\centering
\begin{subfigure}{0.45\textwidth}
\includegraphics[width=1.0\linewidth]{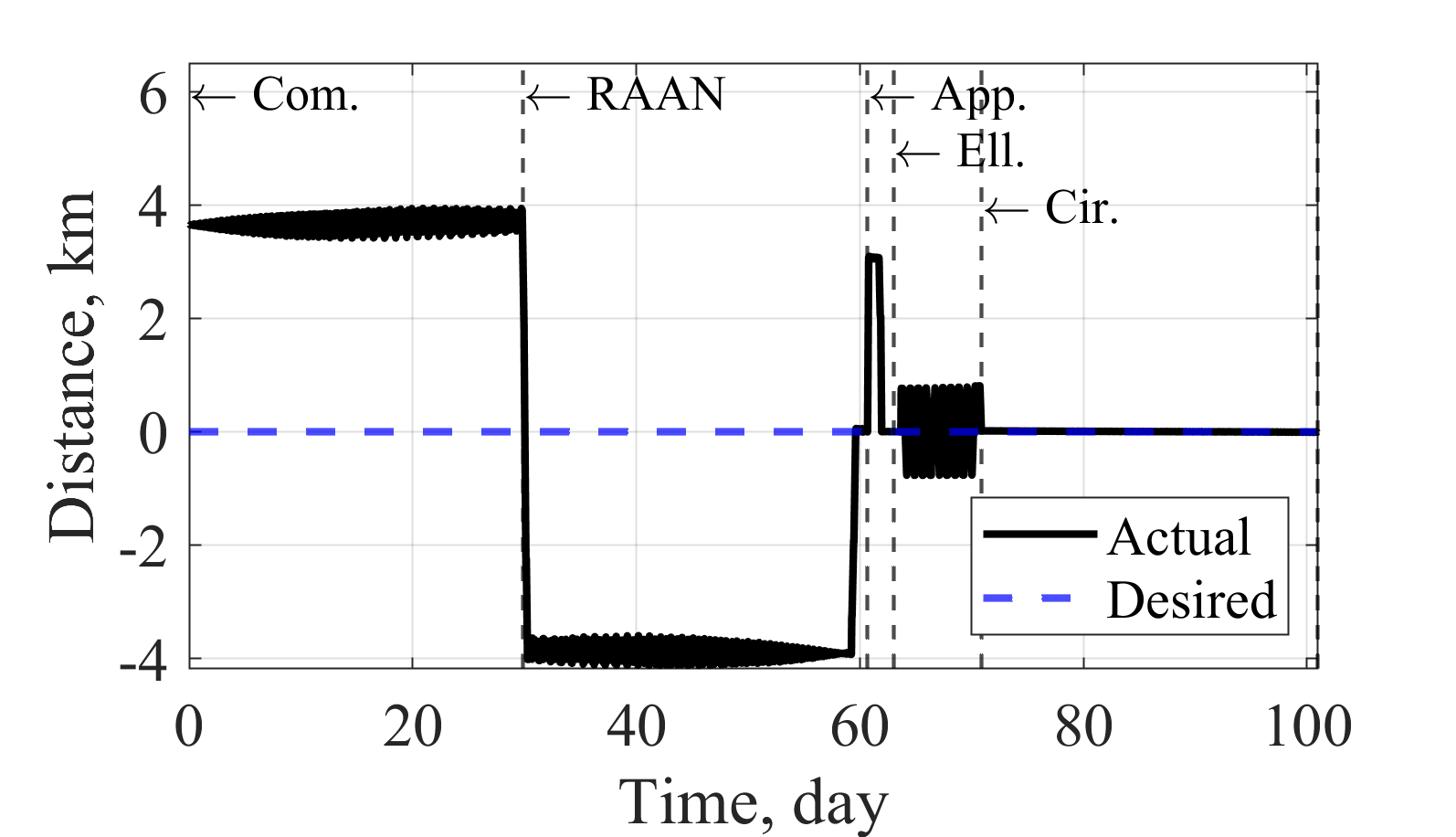}
\caption{$\Delta a$ over time}
\label{fig:delta_a}
\end{subfigure}
\hspace{0cm}
\begin{subfigure}{0.45\textwidth}
\includegraphics[width=1.0\linewidth]{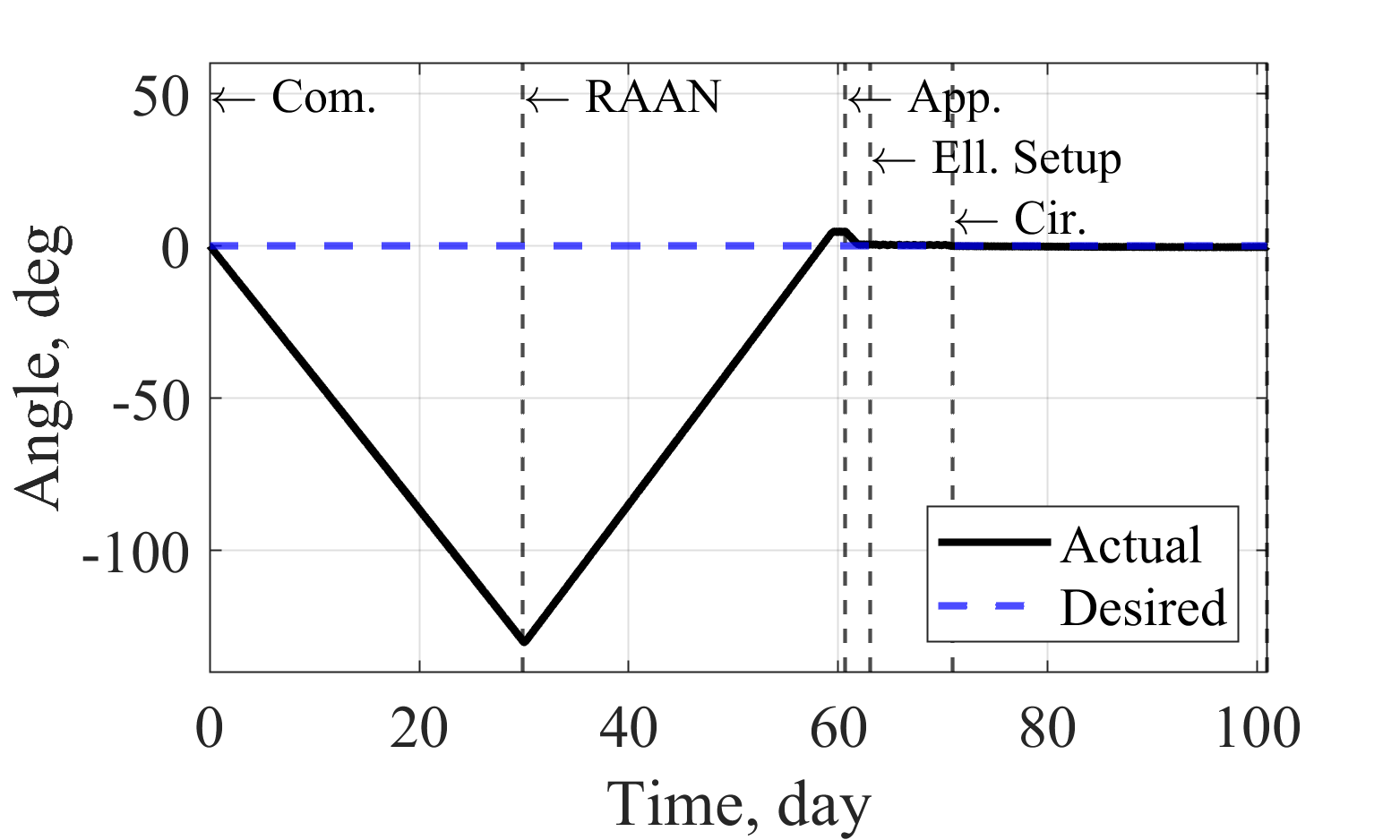}
\caption{$\Delta u$ over time}
\label{fig:delta_u}
\end{subfigure}
\vfill
\begin{subfigure}{0.45\textwidth}
\includegraphics[width=1.0\linewidth]{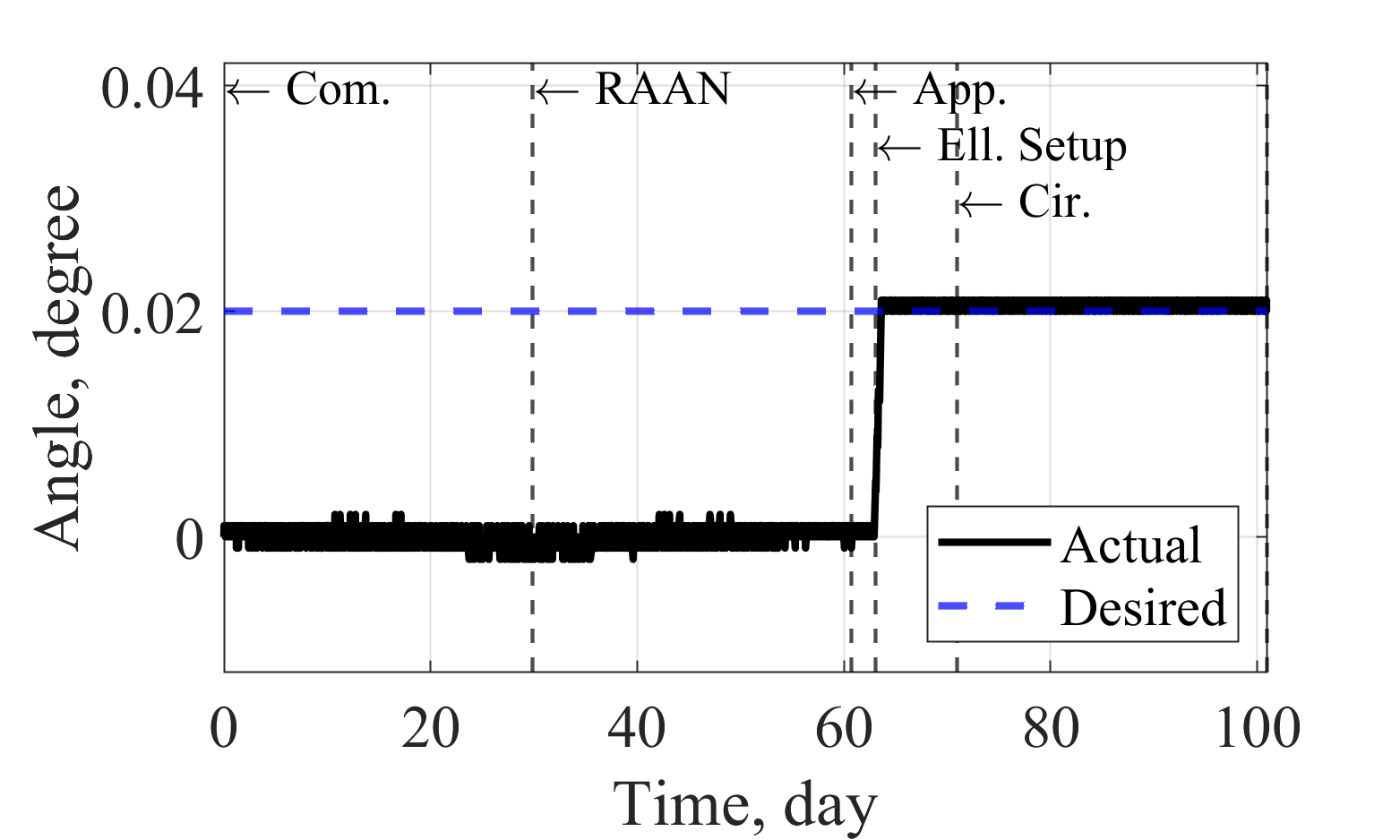}
\caption{$\Delta i$ over time}
\label{fig:delta_i}
\end{subfigure}
\hspace{0cm}
\begin{subfigure}{0.45\textwidth}
\includegraphics[width=1.0\linewidth]{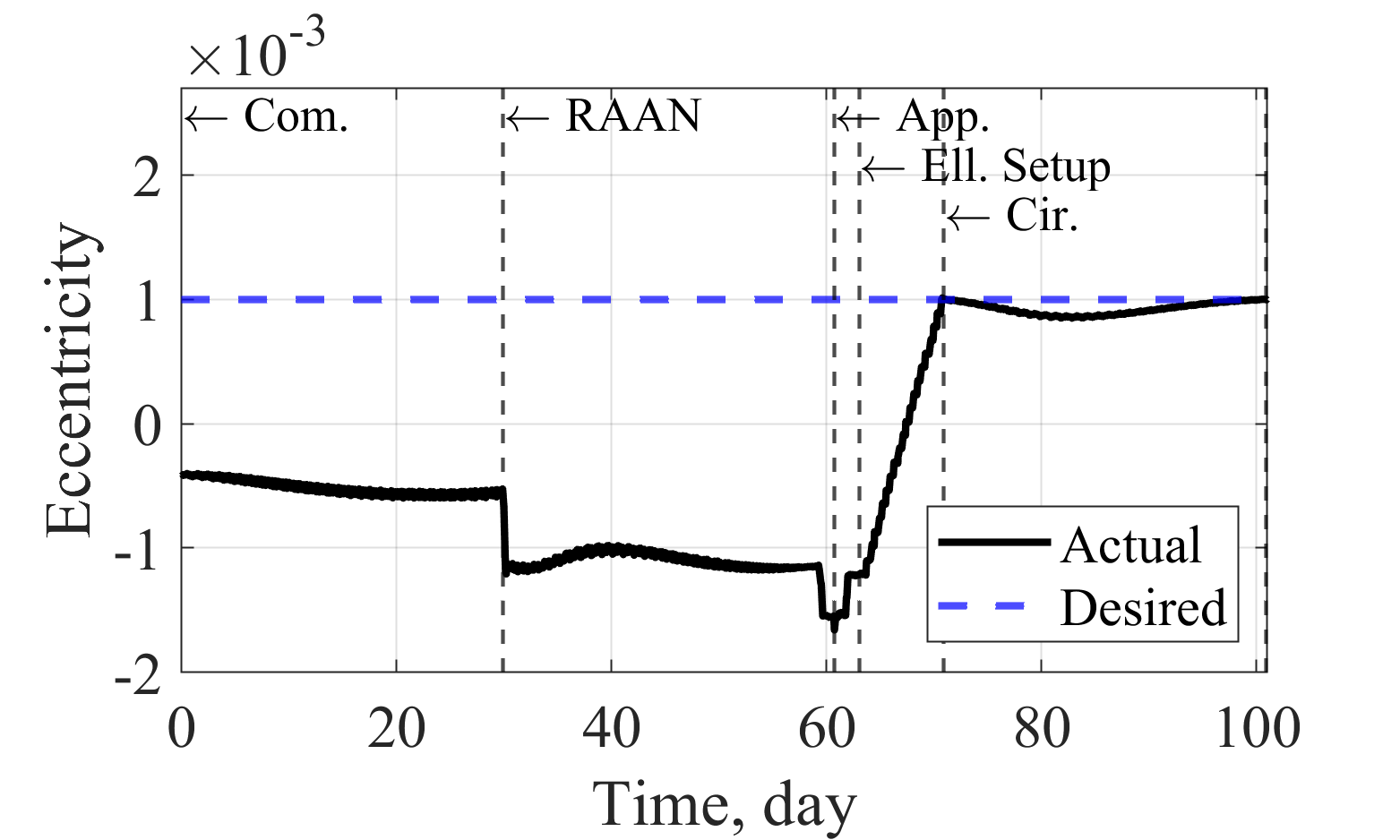}
\caption{$\Delta e$ over time}
\label{fig:delta_e}
\end{subfigure}
\caption{Relative mean orbit elements}
\label{fig:delta_mean_oes}
\end{figure}

Figure~\ref{fig:delta_mean_oes} shows the relative actual and desired mean orbit elements. The $\Delta a$ converges to zero kilometer at the end of the RAAN and approach phases in Figure~\ref{fig:delta_a}. In Figure~\ref{fig:delta_u}, the $\Delta u$ is decreased to 0.5 degree at the end of the approach phase, and remains bounded by 0.5 degree over the ellipse setup phase.
At the end of the ellipse setup phase, the $\Delta i$ and $\Delta e$ converge to the desired values, which is shown in Figure~\ref{fig:delta_i} and Figure~\ref{fig:delta_e}.

\begin{figure}[htb]
\centering
\includegraphics[width=1.0\textwidth]{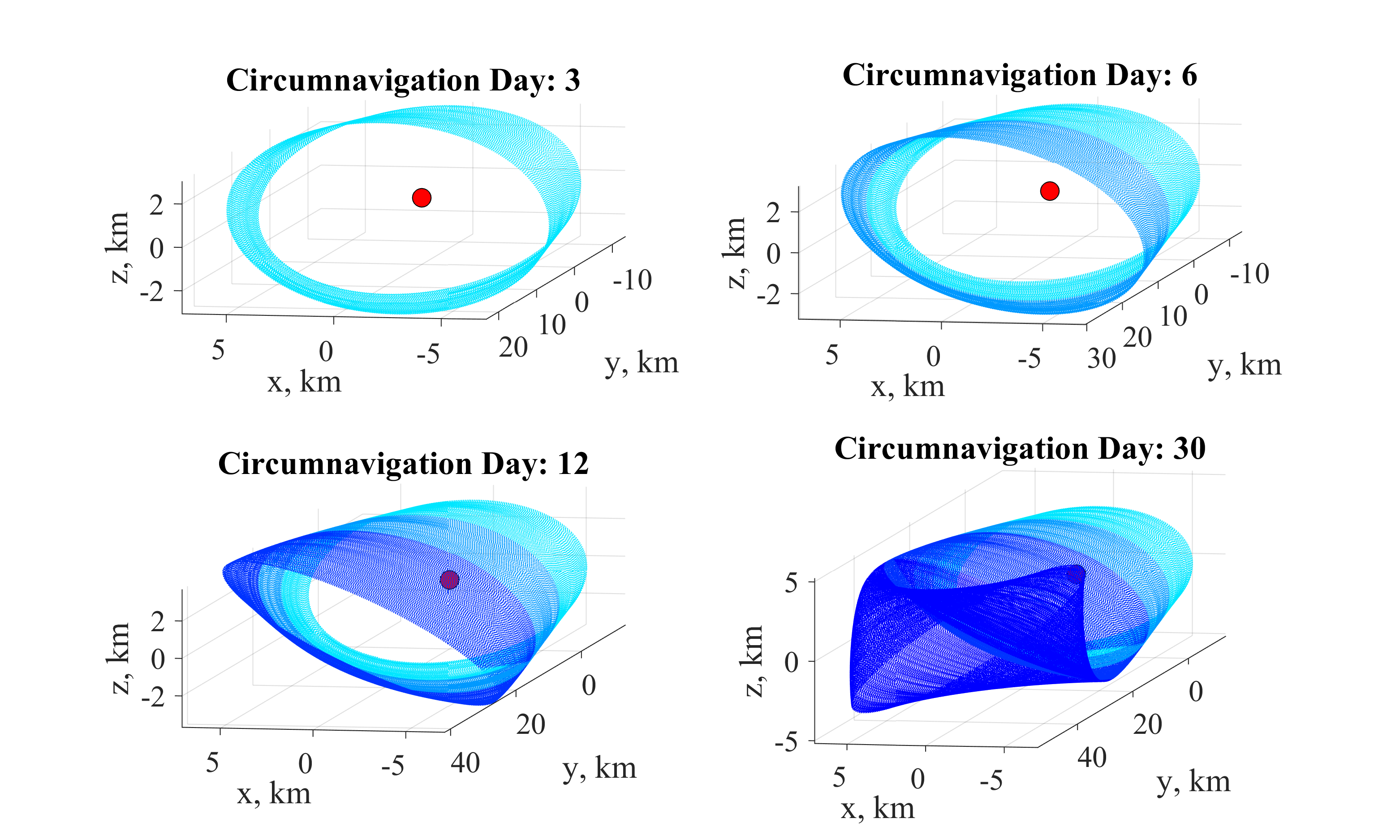} 
\caption{Time evolution of the safety ellipse}
\label{fig:final_SE}
\end{figure}

Note that the $\Delta e$ exhibits variation over the extended circumnavigation in Figure~\ref{fig:delta_e}. This phenomenon can be viewed from the 3D relative trajectory in Figure~\ref{fig:final_SE}. Viewing the safety ellipse from the radial and cross-track plane (x-z plane), the direction of the major axis rotates in the negative y direction, and the magnitude of the minor axis oscillates. In other words, the ellipse is squeezed to two parallel lines, and then expanded to an ellipse periodically. Such time evolution of the safety ellipse is due to the variations of the relative eccentricity/inclination vectors induced by J2 perturbation.\cite{d2006proximity}

\begin{figure}[htb]
\centering
\includegraphics[width=1.0\textwidth]{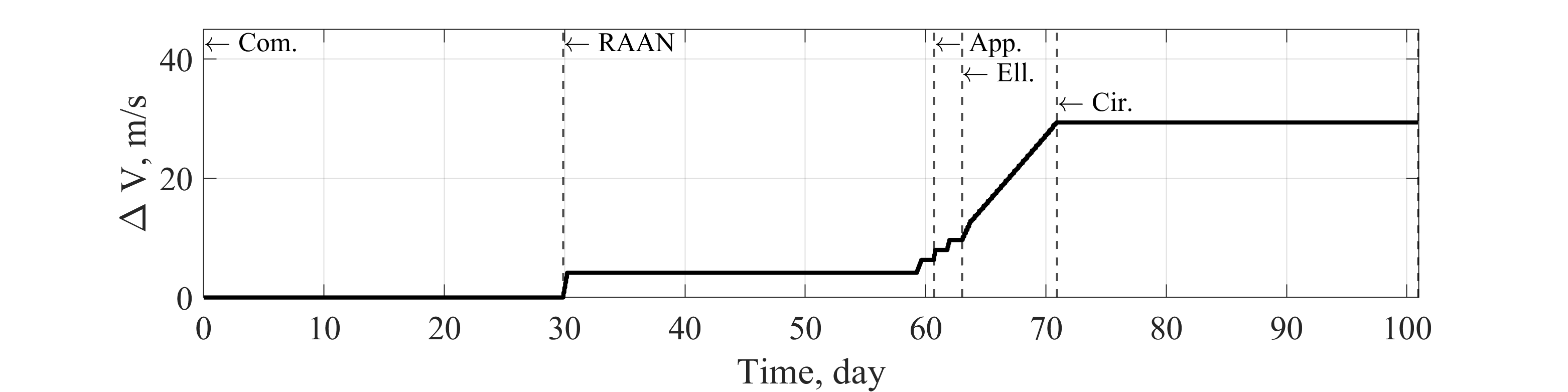} 
\caption{Total $\Delta V$ over time}
\label{fig:final_deltaV}
\end{figure}

\begin{figure}[htb]
\centering
\includegraphics[width=1.0\textwidth]{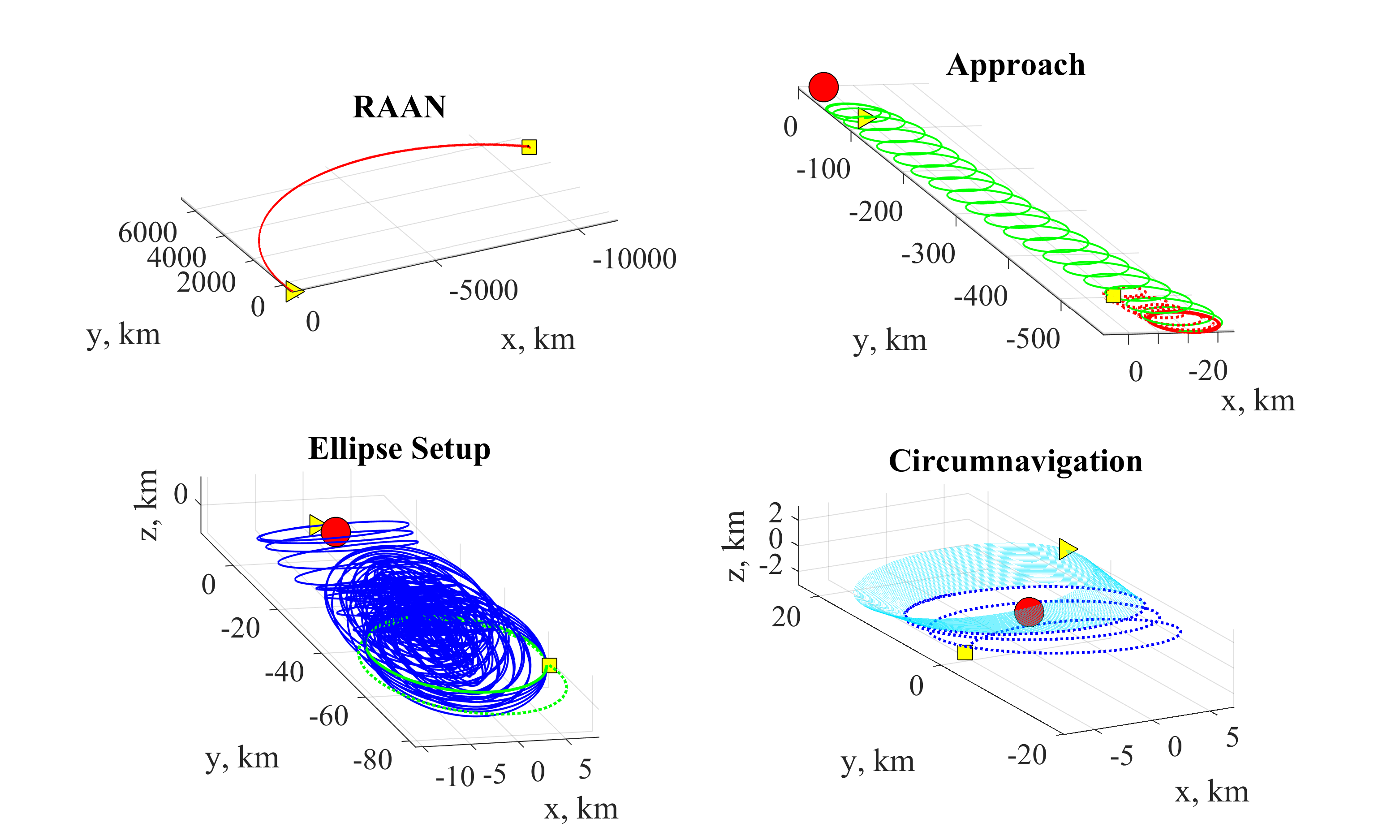} 
\caption{Relative 3D trajectory of the chaser. The red, green, blue, and cyan line indicate the RAAN, Approach, Ellipse Setup, and Circumnavigation phases, respectively. The solid red dot represents the target location, i.e., the origin; the solid yellow square and triangle represent the initial and final locations of each phase.}
\label{fig:final_3DTraj}
\end{figure}

\newpage
The fuel consumption ($\Delta V$) of the overall RPO maneuver is illustrated in Figure~\ref{fig:final_deltaV}. The entire RPO maneuver will require 29.4 m/s $\Delta V$, which accounts for 44$\%$ of the total 67.5 m/s $\Delta V$. Note that the largest fuel consumption occurs in the ellipse setup phase. Lastly, the relative 3D trajectory is plotted in Figure~\ref{fig:final_3DTraj}. At the beginning of the RAAN phase, the relative distance on the x-y plane is around -10,000 km $\times$ 6,000 km. At the end of the RAAN phase, these distances are decreased to -20 km $\times$ -500 km. During the approach phase, the relative distance on the x-y plane is decreased to -10 km $\times$ -60 km. During the ellipse setup phase, the safety ellipse is established (from green ellipse to blue ellipse), and the along-track offset approaches to zero at the end of this phase. The safety ellipse during the nominal circumnavigation phase is plotted. Note that there exists small along-track drift compared to the previous ellipse setup phase, which is an expected result because of the chaser's unforced propagation in the environment with J2 perturbation.\cite{d2006proximity}

\section{Conclusion}
A control framework was developed for the rendezvous and proximity operation (RPO) of two 3U CubeSats. In this control framework, four modularized maneuver block were designed to correct the CubeSat chaser's mean orbit elements while satisfying the constraints imposed by the orbit determination method and the operation of a single electric propulsion thruster. Through a numerical simulation, the CubeSat chaser was shown to approach and circumnavigate the CubeSat target with the implementation of the proposed control framework. After the RPO maneuver, the actual safety ellipse had a three-dimensional size of 14 km $\times$ 27 km $\times$ 8 km, which was similar to the size of the ideal desired safety ellipse: 14 km $\times$ 28 km $\times$ 5 km. In addition, the resulting total fuel consumption of $\Delta V=29.4$ m/s demonstrated the feasibility of performing low Earth orbit RPO with electric propulsion. 

On the other hand, the simulation results exhibited the limitation of the proposed control framework: it does not consider the thruster firings for stabilizing the established safety ellipse. Therefore, developing the control law for RPO missions that require long period of safety ellipse is left for future work.


\bibliographystyle{AAS_publication}   
\bibliography{references}   

\begin{thebibliography}{10}

\bibitem{gangestad2021sat}
J.~W. Gangestad, C.~C. Venturini, D.~A. Hinkley, and G.~Kinum, ``A Sat-to-Sat
  Inspection Demonstration with the AeroCube-10 1.5 U CubeSats,''  2021.

\bibitem{roscoe2018overview}
C.~W. Roscoe, J.~J. Westphal, and E.~Mosleh, ``Overview and GNC design of the
  CubeSat Proximity Operations Demonstration (CPOD) mission,''  {\em Acta
  Astronautica}, Vol.~153, 2018, pp.~410--421.

\bibitem{gaylor2007algorithms}
D.~E. Gaylor and B.~W. Barbee, ``Algorithms for safe spacecraft proximity
  operations,''  {\em AAS/AIAA Spaceflight Mechanics Meeting}, 2007.

\bibitem{navabi2013algebraic}
M.~Navabi, M.~Barati, and H.~Bonyan, ``Algebraic orbit elements difference
  description of dynamics models for satellite formation flying,''  {\em 2013
  6th international conference on recent advances in space technologies
  (RAST)}, IEEE, 2013, pp.~277--280.

\bibitem{shuster2020analytic}
S.~Shuster, {\em Analytic Guidance Strategies for Passively Safe Rendezvous and
  Proximity Operations}.
\newblock PhD thesis, Utah State University, 2020.

\bibitem{garcia2021electric}
A.~Garcia, R.~Linares, Z.~Folcik, P.~Niedfeldt, J.~Gandek, C.~Jewison, and
  P.~Cefola, ``Electric Propulsion Intelligent Control (EPIC) Toolbox for
  Proximity Operations and Safety Analysis in Low-Earth Orbit (LEO),''  2021.

\bibitem{vallado2013fundamentals}
D.~Vallado and W.~McClain, ``Fundamentals of astrodynamics and applications 4th
  Edition,''  2013.

\bibitem{bando2013plane}
M.~Bando and A.~Ichikawa, ``In-Plane Motion Control of
  Hill--Clohessy--Wiltshire Equations by Single Input,''  {\em Journal of
  Guidance, Control, and Dynamics}, Vol.~36, No.~5, 2013, pp.~1512--1522.

\bibitem{willis2017relative}
M.~Willis and S.~D’Amico, ``Relative Spiral Trajectories for Low-Thrust
  Formation Flying,''  {\em 26th International Symposium on Space Flight
  Dynamics, Matsuyama, Japan}, 2017.

\bibitem{schaub2004relative}
H.~Schaub, ``Relative orbit geometry through classical orbit element
  differences,''  {\em Journal of Guidance, Control, and Dynamics}, Vol.~27,
  No.~5, 2004, pp.~839--848.

\bibitem{schaub2000spacecraft}
H.~Schaub, S.~R. Vadali, J.~L. Junkins, and K.~T. Alfriend, ``Spacecraft
  formation flying control using mean orbit elements,''  {\em The Journal of
  the Astronautical Sciences}, Vol.~48, 2000, pp.~69--87.

\bibitem{6581216}
M.~Navabi, M.~Barati, and H.~Bonyan, ``Algebraic orbit elements difference
  description of dynamics models for satellite formation flying,''  {\em 2013
  6th International Conference on Recent Advances in Space Technologies
  (RAST)}, 2013, pp.~277--280, 10.1109/RAST.2013.6581216.

\bibitem{d2006proximity}
S.~D'Amico and O.~Montenbruck, ``Proximity operations of formation-flying
  spacecraft using an eccentricity/inclination vector separation,''  {\em
  Journal of Guidance, Control, and Dynamics}, Vol.~29, No.~3, 2006,
  pp.~554--563.

\end{thebibliography}

\end{document}